\documentclass[aps,prd,amsmath,showpacs]{revtex4}
\usepackage{epsf}
\newcommand{\fr}[2]{\frac{#1}{#2}}
\newcommand{\Ref}[1]{(\ref{#1})}
\newcommand{\be}{\begin{equation}}
\newcommand{\ee}{\end{equation}}
\newcommand{\bn}{\begin{eqnarray}}
\newcommand{\en}{\end{eqnarray}}
\newcommand{\bd}{\begin{displaymath}}
\newcommand{\ed}{\end{displaymath}}
\newcommand{\bnn}{\begin{eqnarray*}}
\newcommand{\enn}{\end{eqnarray*}}
\newcommand{\adb}{\allowdisplaybreaks }
\newcommand{\bs}{\begin{subequations}}
\newcommand{\es}{\end{subequations}}

\begin{document}

\title{Casimir effect in a wormhole spacetime}

\author{Artem R.
Khabibullin${}^{a}$\footnotemark\thanks{e-mail:
arty@theory.kazan-spu.ru}, Nail R.
Khusnutdinov${}^{a}$\footnotemark\thanks{e-mail:
nail@theory.kazan-spu.ru}, Sergey V.
Sushkov${}^{b}$\footnotemark\thanks{e-mail: sushkov@kspu.kcn.ru}}

\address{${}^{a}$Department of Physics,
${}^{b}$Department of Mathematics,\\
Kazan State Pedagogical University, Mezhlauk 1, Kazan 420021,
Russia}

\date{\today}

\begin{abstract}
We consider the Casimir effect for quantized massive scalar field
with non-conformal coupling $\xi$ in a spacetime of wormhole whose
throat is rounded by a spherical shell. In the framework of
zeta-regularization approach we calculate a zero point energy of
scalar field. We found that depending on values of coupling $\xi$,
a mass of field $m$, and/or the throat's radius $a$ the Casimir
force may be both attractive and repulsive, and even equals to
zero.
\end{abstract}

\maketitle
\section{Introduction}\label{Intro}

The central problem of wormhole physics consists of the fact that
wormholes are accompanied by unavoidable violations of the null
energy condition, i.e., the matter threading the wormhole's throat
has to possess ``exotic'' properties. The classical matter does
satisfy the usual energy conditions, hence wormholes cannot arise
as solutions of classical relativity and matter. If they exist,
they must belong to the realm of semiclassical or perhaps quantum
gravity. In the absence of the complete theory of quantum gravity,
the semiclassical approach begins to play the most important role
for examining wormholes. Recently the self-consistent wormholes in
the semiclassical gravity were studied numerically in Refs
\cite{HocPopSus97,KhuSus02,Khu03,Gar05}. It was shown that the
semiclassical Einstein equations provide an existence of wormholes
supported by energy of vacuum fluctuations. However, it should be
stressed that a natural size of semiclassical vacuum wormholes
(say, a radius of wormhole's throat $a$) should be of Planckian
scales or less. This fact can be easily argued by simple
dimensional considerations \cite{ForRom96}. In order to obtain
semiclassical wormholes having scales larger than Planckian one
has to consider either non-vacuum states of quantized fields (say,
thermal states with a temperature $T>0$) or a vacuum polarization
(the Casimir effect) which may happen due to some external
boundaries (with a typical scale $R$) existing in a wormhole
spacetime. In the both cases there appears an additional
dimensional macroscopical parameter (say $R$) which may result in
enlargement of wormhole's size.

In this paper we will study the Casimir effect in a wormhole
spacetime. For this aim we will consider a static spherically
symmetric wormhole joining two different universes (asymptotically
flat regions). We will also suppose that each universe contains a
perfectly conducting spherical shell rounding the throat. These
shells will dictate the Dirichlet boundary conditions for a
physical field and, as the result, produce a vacuum polarization.
Note that this problem is closely related to the known problem
which was investigated by Boyer \cite{Boy68} who studied the
Casimir effect of a perfectly conducting sphere in Minkowski
spacetime (see also \cite{BorEliKirLes97}). However, there is an
essential difference which is expressed in different topologies of
wormhole and Minkowski spacetimes. A semitransparent sphere as
well as semitransparent boundary condition were investigated in
Refs.
\cite{BorVas99,Sca99,Sca00,BorVas04,GraJafKheQuaShrWei04,Mil04}.
The consideration of the delta-like potential which models a
semitransparent boundary condition in quantum field theory cause
some problems and there is ambiguity in renormalization procedure
(see the Refs. \cite{BorVas04,GraJafKheQuaShrWei04,Mil04} and
references therein). Thermal corrections to the one-loop effective
action on singular potential background was considered recently in
Ref. \cite{MckNay05}.

We will adopt a simple geometrical model of wormhole spacetime:
the short-throat flat-space wormhole which was suggested and
exploited in Ref. \cite{KhuSus02}. The model represents two
identical copies of Minkowski spacetime; from each copy a
spherical region is excised, and then boundaries of those regions
are to be identified. The spacetime of the model is everywhere
flat except a throat, i.e., a two-dimensional singular spherical
surface. We will assume that the wormhole's throat is rounding by
two perfectly conducting spherical shells (in each copy of
Minkowski spacetime) and calculate the zero-point energy of a
massive scalar field on this background. In the end of
calculations the radius of one sphere will tend to infinity giving
the Casimir energy for single sphere. For calculations we will use
the zeta function regularization approach \cite{DowCri76,ZetaBook}
which was developed in Refs.
\cite{Method,BorEliKirLes97,KhuBor99,BezBezKhu01,BorMohMos01}. In
framework of this approach, the ground state energy of scalar
field $\phi$ is given by
\be
E(s) = \fr 12 \mu^{2s} \zeta_{\cal L} \left(s - \fr 12\right),
\label{DefZeta}
\ee
where \bd \zeta_{\cal L} (s) = \sum_{(n)} \left(\lambda_{(n)}^2 +
m^2 \right)^{-s} \ed is the zeta function of the corresponding
Laplace operator. The parameter $\mu$, having the dimension of
mass, makes right the dimension of regularized energy.
The $\lambda_{(n)}^2$ are eigenvalues of the
three dimensional Laplace operator ${\cal L} = \triangle - \xi
\mathcal{R}$
\be
(\triangle - \xi \mathcal{R}) \phi_{(n)} = \lambda_{(n)}^2
\phi_{(n)},\label{eigenvalue}
\ee
where $\mathcal{R}$ is the curvature scalar (which is singular in
framework of our model, see Eq. \Ref{singcurv}).

The expression \Ref{DefZeta} is divergent in the limit $s\to 0$
which we are interested in. For renormalization we subtract from
\Ref{DefZeta} the divergent part of it
\be
E^{\rm ren} = \lim_{s\to 0} \left(E(s) - E^{\rm div} (s)\right),
\label{ERen}
\ee
where
\bd
E^{\rm div} (s) = \lim_{m\to \infty} E(s).
\ed
By virtue of the heat kernel expansion of zeta function is the
asymptotic expansion for large mass, the divergent part has the
following form (in $3+1$ dimensions)
\bn
E^{\rm div} (s) &=& \fr 12\left(\fr\mu m\right)^{2s} \fr
1{(4\pi)^{3/2} \Gamma (s - \fr 12)} \label{div_terms} \\
&\times& \left\{ B_0 m^4\Gamma (s-2) + B_{1/2}m^3\Gamma (s-\fr 32)
+ B_1 m^2 \Gamma (s-1) + B_{3/2} m \Gamma (s-\fr 12) + B_2 \Gamma
(s) \right\}, \nonumber
\en
where $B_\alpha$ are the heat kernel coefficients of operator
${\cal L}$. In the case of singular potential (singular scalar
curvature) one has to use specific formulae from Refs.
\cite{BorVas99,GilKirVas01} for calculation the heat kernel
coefficients (see also a recent review \cite{Vas02}).

Finally, the renormalized ground state energy \Ref{ERen} should
obey the normalization condition
\bd
\lim_{m\to \infty} E^{\rm
ren} = 0.
\ed For more details of approach see review
\cite{BorMohMos01}.

The organization of the paper is the following. In Sec.
\ref{model} we describe a spacetime of wormhole in the
short-throat flat-space approximation. In Sec. \ref{ZPE} we
analyze the solution of equation of motion for massive scalar
field and obtain close expression for zero point energy.
In Sec. \ref{conclusion} we discuss obtained results and make some
speculations.

We use units $\hbar = c = G = 1$. The signature of the spacetime,
the sign of the Riemann and Ricci tensors, is the same as in the
book by Hawking and Ellis \cite{HawEllBook}.

\section{The geometry of the model}\label{model}

We will take a metric of static spherically symmetric wormhole in
a simple form:
\be\label{Metric}
ds^2 = -dt^2 + d\rho^2 + r^2(\rho)( d\theta^2 + \sin^2\theta
d\varphi^2),
\ee
where $\rho$ is a proper radial distance, $\rho \in
(-\infty,\infty)$. The function $r(\rho)$ describes the profile of
throat. In the paper we adopt the model suggested in the Ref.
\cite{KhuSus02} which was called there as short-throat flat-space
approximation. In framework of this model the shape function
$r(\rho)$ is
\bd
r(\rho)=|\rho\,|+a,
\ed
with $a>0$. $r(\rho)$ is always positive and has the minimum at
$\rho = 0$: $r(0)=a$, where $a$ is a radius of throat. It is easy
to see that in two regions ${\cal D}_+\!:\,\rho>0$ and ${\cal D}_-
\!:\,\rho<0$ one can introduce new radial coordinates
$r_\pm=\pm\rho+a$, respectively, and rewrite the metric
\Ref{Metric} in the usual spherical coordinates:
\bd
ds^2=-dt^2+dr_\pm^2+r_\pm^2(d\theta^2+\sin^2\theta\,d\varphi^2 ),
\ed
This form of the metric explicitly indicates that the regions
${\cal D}_+$ and ${\cal D}_-$ are flat. However, note that such
the change of coordinates $r_\pm=\pm\rho+a$ is not global, because
it is ill defined at the throat $\rho=0$. Hence, as was expected,
the spacetime is curved at the wormhole throat. To illustrate this
we calculate the Ricci tensor in the metric \Ref{Metric}:
\bn
\mathcal{R}^\rho_\rho &=& -\fr{2r''}{r} = -4\fr{\delta (\rho)}{a}, \nonumber\\
\mathcal{R}^\theta_\theta = \mathcal{R}^\varphi_\varphi &=&
-\fr{-1+r'^2 + r r''}{r^2} = -2\fr{\delta (\rho)}{a},\label{singcurv}\\
\mathcal{R} &=& -\fr{2(-1+r'^2 + 2r r'')}{r^2} = -8\fr{\delta
(\rho)}{a}.\nonumber
\en
The energy-momentum tensor corresponding to this metric has the
diagonal form from which we observe that the source of this metric
possesses the following energy density and pressure:
\bnn
\varepsilon &=& -\frac{-1 + r'^2 + 2rr''}{8\pi r^2} =
-\fr{\delta (\rho)}{2\pi a},\adb\\
p_\rho &=&\frac{-1 + r'^2}{8\pi r^2} = 0,\adb\\
p_\theta &=& p_\varphi = \frac{r''}{8\pi r} = \fr{\delta
(\rho)}{4\pi a}.
\enn

\section{Zero point energy}\label{ZPE}

Let us now consider a scalar field $\phi$ in the spacetime with
the metric \Ref{Metric}. The equation for eigenvalues of operator
${\cal L}$ is
\be\label{eqmotion}
(\triangle - \xi {\cal R}) \phi_{(n)} = \lambda_{(n)}^2
\phi_{(n)},
\ee
where ${\cal R}$ is the scalar curvature, $\xi$ is an arbitrary
coupling with ${\cal R}$ and $\triangle =
g^{\alpha\beta}\nabla_\alpha \nabla_\beta$, $\alpha=1,2,3$. Due to
the spherical symmetry of spacetime \Ref{Metric}, a general
solution to the equation \Ref{eqmotion} can be found in the
following form:
\bd \phi(\rho,\theta,\varphi) = u(\rho)Y_{ln}(\theta,\varphi), \ed
where $Y_{ln}(\theta,\varphi)$ are spherical functions,
$l=0,1,2,\dots$, $n=0,\pm1,\pm2,\dots,\pm l$, and a function
$u(\rho)$ obeys the radial equation
\be\label{radialeq}
u''+2\frac{r'}{r}u'+\left(\lambda^2-\frac{l(l+1)}{r^2} - \xi {\cal
R}\right)u=0,
\ee
where a prime denotes the derivative with respect $\rho$,
$\lambda=\sqrt{\omega^2-m^2}$ and scalar curvature ${\cal R} = -
8\delta(\rho)/a$. For new function $w = u r$ this equation reads
\bd
w''+\left(\lambda^2-\frac{l(l+1)}{r^2} - \xi {\cal R} -
\fr{r''}{r}\right)w=0,
\ed
and looks like the Schr\"odinger equation for massive particle
with mass $M$ with total energy $E=\lambda^2/2M$ and potential
energy
\be\label{potential}
U=(\xi {\cal R} + \fr{r''}{r})/2M = \fr{1-4\xi}{aM} \delta(\rho).
\ee
Therefore, the $\xi > 1/4$ corresponds to negative potential.

Unfortunately, in our case it is impossible to find in manifest
form the spectrum of operator ${\cal L}$ given by Eq.
\Ref{eqmotion}. For this reason, we will use an approach developed
in Refs.
\cite{Method,BorEliKirLes97,KhuBor99,BorMohMos01,BezBezKhu01}.
This approach does not need an explicit form of spectrum. The
spectrum of an operator is usually found from some boundary
conditions which look like an equation $\Psi(\lambda)=0$ where
function $\Psi$ consists of the solutions of Eq. \Ref{radialeq}
and depends additionally on other parameters of problem. It was
shown in Refs.
\cite{Method,BorEliKirLes97,KhuBor99,BorMohMos01,BezBezKhu01} that
the zero point energy may be represented in the following form:
\be\label{mainmain}
E(s) = -\mu^{2s}\fr{\cos (\pi s)}{2\pi} \sum_{(n)}d_n\int_m^\infty
dk (k^2-m^2)^{1/2 -s} \fr{\partial}{\partial k}\ln \Psi(ik),
\ee
with the function $\Psi$ taken on the imaginary axes. The sum is
taken over all numbers of problem and $d_n$ is degenerate of state
\footnote{For the spherical symmetry case $(n) = l$ and $d_n = 2
l+1 = 2\nu$.}. This formula takes into account the possible
boundary states, too. If they exist we have to include them
additively at the beginning in the Eq. \Ref{DefZeta}. But
integration over interval $|k| < m$ (the possible boundary states
exist in this domain) will cancel this contribution. For this
reason the integration in the formula \Ref{mainmain} is started
from the energy $k=m$. Therefore, hereinafter we will consider the
solution of the Eq. \Ref{radialeq} for negative energy that is in
imaginary axes $\lambda = i k$. The main problem is now reduced to
finding the function $\Psi$. Thus, now we need no explicit form of
spectrum of operator ${\cal L}$.

In the flat regions ${\cal D}_\pm$, where $r(\rho)=\pm\rho+a$,
$r'(\rho)=\pm 1$, ${\cal R}(\rho)=0$, and in imaginary axes the
Eq.\Ref{radialeq} reads
\be\label{radialeqimag}
u''+\frac{2}{\rho\pm a}u' - \left(k^2 + \frac{l(l+1)}{(\rho\pm
a)^2}\right)u=0.
\ee

A general solution of this equation can be written as
\be\label{sol}
u^{\pm}[k(a\pm \rho)] = A^{\pm} \sqrt{\fr{\pi}{2 k (a\pm \rho)}}
I_\nu [k (a\pm \rho)] + B^{\pm} \sqrt{\fr{\pi}{2 k (a\pm \rho)}}
K_\nu [k (a\pm \rho)],
\ee
where $I_\nu, K_\nu$ are the Bessel functions of second kind, $\nu =
l+1/2$, and $A^{\pm}$, $B^{\pm}$ are four arbitrary constants.

The solutions $u^{\pm}[k (\rho\pm a)]$ have been obtained in the
flat regions ${\cal D}_\pm$ separately. To find a solution in the
whole spacetime we must impose matching conditions for $u^{\pm}[k
(\rho\pm a)]$ at the throat $\rho=0$. The first condition demands
that the solution has to be continuous at $\rho=0$. This gives
\bs\label{cond}
\be\label{cond1}
u^{-} [ka]=u^{+} [ka].
\ee
To obtain the second condition we integrate Eq.\Ref{radialeq}
within the interval $(-\epsilon,\epsilon)$ and then go to the
limit $\epsilon\to0$.  It gives the second condition
\be\label{cond2}
\left.- \frac{du^{-}[x]}{dx}\right|_{x= ka} = \left.
\frac{du^{+}[x]}{dx} \right|_{x= k a} + \frac{8\xi}{ka} u^{+}[ka].
\ee
Therefore, the general solution of Eq. \Ref{radialeqimag} depends
on two constants, only. Two other constants may be found from Eqs.
\Ref{cond1} and \Ref{cond2}.

In addition to two matching conditions \Ref{cond1} and \Ref{cond2}
we impose two boundary conditions. We round the wormhole throat by
sphere of radius $a+R$ ($\rho = R$) in region ${\cal D}_+$, and by
sphere of radius $a+R'$ ($\rho = - R'$) in region ${\cal D}_-$.
Therefore the space of wormhole is divided by two spheres to three
regions: the space of finite volume between spheres and two
infinite volume spaces out of spheres. We suppose that the scalar
field obeys the Dirichlet boundary condition on both of these
spheres which means the perfect conductivity of spheres:
\bn
u^{-} [k(R'+a)] &=& 0, \label{cond3}\\ u^{+} [k(R+a)] &=&
0.\label{cond4}
\en
\es

The four conditions \Ref{cond} obtained represent a homogeneous
system of linear algebraic equations for four coefficients
$A^{\pm}$, $B^{\pm}$. As is known, such a system has a nontrivial
solution if and only if the matrix of coefficients is degenerate.
Hence we get
\be\label{det} \left|
\begin{array}{cccc}
-I_\nu[ka] & -K_\nu[ka] & I_\nu[ka] & K_\nu[ka] \\
I'_\nu[ka] + \fr{16\xi -1}{2ka}I_\nu[ka] & K'_\nu[ka] + \fr{16\xi
-1}{2ka}K_\nu[ka] & I'_\nu[ka] - \fr{1}{2ka}I_\nu[ka] &
K'_\nu[ka] - \fr{1}{2ka}K_\nu[ka] \\
I_\nu[k(a+R)] & K_\nu[k(a+R)] & 0 & 0 \\
0 & 0 & I_\nu[k(a+R')] & K_\nu[k(a+R')]
\end{array}
\right|=0.
\ee
\bs
After some algebra the above formula can be reduced to the
following relation for function $\Psi$ which we need for
calculation of the energy \Ref{mainmain}:
\bn\label{spectr1} \Psi_{in} &=&
I_\nu[k(a+R')] \left(\Psi^* \left[\left(\xi - \fr 18\right)
K_\nu[ka] + \fr{ka}{4}K'_\nu[ka]\right] - \fr
18K_\nu[k(a+R)]  \right) \\
&-& K_\nu[k(a+R')] \left(\Psi^* \left[\left(\xi - \fr 18\right)
I_\nu[ka] + \fr{ka}{4}I'_\nu[ka]\right] - \fr 18I_\nu[k(a+R)]
\right)=0, \nonumber
\en
with
\bd \Psi^*= I_\nu[k(a+R)]K_\nu[ka] - K_\nu[k(a+R)]
I_\nu[ka].
\ed
In the case $R'=R$ above expression coincides with that obtained
in Ref. \cite{KhuSus02}. In this case $\Psi_{in}$ may be
represented as follows: $\Psi_{in}= \Psi^1_l\Psi^2_l$, where
\bnn
\Psi^1_l &=& \Psi^* = I_\nu[k(a+R)]K_\nu[ka] - K_\nu[k(a+R)]
I_\nu[ka],\\
\Psi^2_l &=& \left(\xi - \fr 18\right) \Psi^* +
\fr{ka}{4}\left[I_\nu[k(a+R)]K'_\nu[ka] - K_\nu[k(a+R)]
I'_\nu[ka]\right].
\enn

The solutions of Eq. \Ref{spectr1} gives the spectrum of energies
between the spheres $R$ and $R'$.
The spectra for regions out of these spheres can be found as
follows:
\bn
\Psi_{out}^1 = K_\nu[k(a+R)], \label{outside}\\
\Psi_{out}^2 = K_\nu[k(a+R')].
\en
\es
Indeed, let us consider the energy spectrum of field in space
between two spheres with radii $R$ and $\widetilde{R} > R$ and
Dirichlet boundary conditions on them. The solution is a linear
combination of two modified Bessel functions
\bd
u_{R\widetilde{R}} = C_1 I_\nu[k\rho] + C_2K_\nu[k\rho].
\ed
The Dirichlet boundary conditions give two equations
\bnn
C_1 I_\nu[k(a+R)] + C_2K_\nu[k(a+R)] = 0, \\
C_1 I_\nu[k(a+\widetilde{R})] + C_2K_\nu[k(a+\widetilde{R})] = 0.
\enn
Using these equations we may represent the solution in the
following form:
\bd
u_{R\widetilde{R}} = \fr{C_1}{K_\nu[k(a+R)]}
\left\{I_\nu[k\rho]
\fr{K_\nu[k(a+\widetilde{R})]}{I_\nu[k(a+\widetilde{R})]} -
K_\nu[k\rho]\right\}.
\ed
Let us now assume that $\widetilde{R}\to \infty$.
In this limit the solution takes the following form:
\bd
u_{R\infty} = C K_\nu[k\rho].
\ed
The Dirichlet boundary condition for this solution on the sphere
of radius $R$ gives the equation \Ref{outside}. As expected this
condition coincides with expression for space out of sphere of
radius $a+R$ in Minkowski spacetime \cite{BorEliKirLes97}. It is
obviously because the spacetime out of sphere (in general out of
throat) is exactly Minkowski spacetime.

Therefore the regularized total energy \Ref{mainmain} reads
\be\label{main}
E(s) = -\mu^{2s}\fr{\cos (\pi s)}{\pi}\sum_{l=0}^\infty \nu
\int_m^\infty dk (k^2-m^2)^{1/2 -s} \fr{\partial}{\partial
k}\left[ \ln \Psi_{in} + \ln \Psi_{out}^1 + \ln
\Psi_{out}^2\right].
\ee
Regrouping terms we can rewrite the above formula in the form
having clear physical sense of each term:
\be E(s) = \triangle
E(s) + E_{R}^M(s) + E_{R'}^M(s), \label{Es}
\ee
where
\bn E_R^M (s) &=& -\mu^{2s}\fr{\cos (\pi s)}{\pi}
\sum_{l=0}^\infty \nu \int_m^\infty dk (k^2-m^2)^{1/2 -s}
\fr{\partial}{\partial k} \ln
I_\nu[k(a+R)]K_\nu[k(a+R)],\label{ER}\\
E_{R'}^M (s) &=& -\mu^{2s}\fr{\cos (\pi s)}{\pi} \sum_{l=0}^\infty
\nu \int_m^\infty dk (k^2-m^2)^{1/2 -s} \fr{\partial}{\partial k}
\ln I_\nu[k(a+R')]K_\nu[k(a+R')],\label{ER'}\\
\triangle E(s) &=& -\mu^{2s}\fr{\cos (\pi s)}{\pi}
\sum_{l=0}^\infty \nu \int_m^\infty dk (k^2-m^2)^{1/2 -s}
\fr{\partial}{\partial k} \ln \Psi\label{deltaE}
\en
and \bd \Psi = \fr{\Psi_{in}}{I_\nu[k(a+R')]I_\nu[k(a+R)]} \ed The
term $E_{R}^M(s)$ in the formula \Ref{Es} is nothing but a zero
point energy of sphere of radius $a+R$ in Minkowski spacetime with
Dirichret boundary condition on the sphere \cite{BorEliKirLes97};
note that the term $E_{R'}^M(s)$ has an analogous sense.

Now we are ready to calculate the Casimir energy for two spherical
boundaries by using expression \Ref{main} and Eq. \Ref{ERen}. Then
let us consider the Boyer's problem. We consider "gedanken
experiment": we take a single conducting sphere and measure the
Casimir force in this situation. For this reason we have to take a
limit $R' \to \infty$. In this case the energy \Ref{ER'} tends to
zero, and so the term $\triangle E(s)$ in Eq. \Ref{deltaE}
represents the difference between Casimir energies of a sphere
rounding the wormhole and a sphere of the same radius in Minkowski
spacetime without wormhole. In the limit $R'\to\infty$ we find
\be\label{psiR}
\Psi = \left( K_\nu[ka] - I_\nu[ka]
\fr{K_\nu[k(a+R)]}{I_\nu[k(a+R)]}\right) \left(\left(\xi- \fr
18\right) K_\nu[ka] + \fr{ka}{4} K'_\nu[ka]\right) - \fr 18
\fr{K_\nu[k(a+R)]}{I_\nu[k(a+R)]}
\ee
If one turns $R\to \infty$ then the energy $E^M_R$ tends to zero
and so
\be\label{psiinfty}
\Psi \to K_\nu[ka]\left(\left(\xi- \fr 18\right) K_\nu[ka] +
\fr{ka}{4} K'_\nu[ka]\right).
\ee
This expression  coincides exactly with that obtained in Ref.
\cite{KhuSus02} and describes the zero point energy for whole
wormhole spacetime without any additional spherical shells.

A comment is in order. As already noted the positive $\xi$
corresponds to attractive potential and therefore the boundary
states may appear. The appearance of boundary states with
delta-like potential has been observed in Ref. \cite{MamTru82}.
Thus, we have to take into account the boundary states at the
beginning. Nevertheless, the final formula \Ref{main} contains
these boundary states, as it was noted in Ref. \cite{Method}. But
it is necessary to note, that in this paper we will consider $\xi
< 1/4$. Indeed, let us consider for example $l=0$. In this case
\bd
\Psi = \fr\pi{8} e^{-ka} \cosh (k(a+R)) \left\{ \cosh (kR) +
\left[2 \fr{1 - 4\xi}{ka} +1 \right]\sinh (kR)\right\}.
\ed
For $\xi >1/4$ this expression may be equal to zero for some value
of $k>m$, $R$ and $a$ and integral \Ref{main} will be divergent.
As noted in Ref. \cite{MamTru82} in this case we can not  use the
present theory. The same boundary for $\xi$ was noted in Ref.
\cite{KhuSus02}. This statement is easy to see from expression for
potential energy given by Eq. \Ref{potential}. For $\xi >1/4$ the
energy is negative and the boundary states may appear.

The general strategy of the subsequent calculations is following
(for more details see Refs.
\cite{Method,BorEliKirLes97,KhuBor99,BorMohMos01,BezBezKhu01}). To
single out in manifest form the divergent part of regularized
energy we subtract from and add to integrand in Eq. \Ref{main} its
uniform expansion over $1/\nu$. It is obviously that it is enough
to subtract expansion up to $1/\nu^2$, the next term will give the
converge series. We may set $s=0$ in the part from which we had
subtracted the uniform expansion because it is now finite (see Eq.
\Ref{A}). The divergent singled out part will contain the standard
divergent terms given by Eq. \Ref{div_terms} and some finite terms
which we calculate in manifest form (all terms except $A$ in
\Ref{dE}).

The uniform asymptotic expansions both \Ref{psiR} and
\Ref{psiinfty} are the same for $R\not =0$. Indeed, in this case
the ratios
\bnn
\fr{I_\nu[ka]}{K_\nu[ka]}\fr{K_\nu[k(a+R)]}{I_\nu[k(a+R)]}
&\approx& e^{- 2 \nu \ln (1+\fr Ra)},\\
\fr{1}{K^2_\nu[ka]}\fr{K_\nu[k(a+R)]}{I_\nu[k(a+R)]} &\approx&
2\nu e^{- 2 \nu \ln (1+\fr Ra)}
\enn
are exponentially small and we may neglect them. The well-known
uniform expansions of Bessel functions \cite{AbrSte} were used in
these expressions. For this reason we may disregard this fraction
in Eq. \Ref{psiR} and arrive to Eq. \Ref{psiinfty}. This is a key
observation for next calculations. Due to this observation the
divergent part which we have to subtract for renormalization from
\Ref{deltaE} has been already calculated in Ref. \cite{KhuSus02}.
By using the results of this paper we may write out the expression
for renormalized zero point energy:
\bn
\triangle E &=& -\fr 1{32 \pi^2 a} \left(b \ln \beta^2 +
\Omega\right), \label{dE}\\
\Omega &=& A + \sum_{k=-1}^3 \omega_k (\beta),\\
b &=& \fr 12 b_0 \beta^4 - b_1 \beta^2 + b_2,
\en
where
\bn
A &=& 32\pi \sum_{l=0}^\infty \nu^2 \int_{\beta/\nu}^\infty dy
\sqrt{y^2 - \fr{\beta^2}{\nu^2}} \fr{\partial}{\partial y}
\left(\ln\Psi + 2\nu \eta (y) + \fr 1\nu N_1 - \fr 1{\nu^2} N_2 +
\fr 1{\nu^3} N_3\right),\label{A}\\
\Psi &=& \left( K_\nu[\nu y] - I_\nu[\nu y] \fr{K_\nu[\nu
y(1+x)]}{I_\nu[\nu y(1+x)]}\right) \left(\left(\xi- \fr 18\right)
K_\nu[\nu y] + \fr{\nu y}{4} K'_\nu[\nu y]\right) - \fr 18
\fr{K_\nu[\nu y(1+x)]}{I_\nu[\nu y(1+x)]},
\en
$b_k$ are the heat kernel coefficients, $\beta = ma$ is a
dimensionless parameter of mass, and $x=R/a$ is a dimensionless
parameter of sphere's radius. The explicit form of heat kernel
coefficients $b_k$, and also expressions for $\omega_k,\ N_k,\
\eta$ are given in the Ref. \cite{KhuSus02}. Note that they do not
depend on the radius of sphere $R$. The only dependence on $R$ is
contained in the coefficient $A$ which has to be calculated
numerically. The expression for contribution of the sphere in
Minkowski spacetime \Ref{ER} may be found in Ref.
\cite{BorEliKirLes97}. We only have to make a change $R\to a+R$.

\begin{figure}
\begin{center}
\epsfxsize=5.5truecm\epsfbox{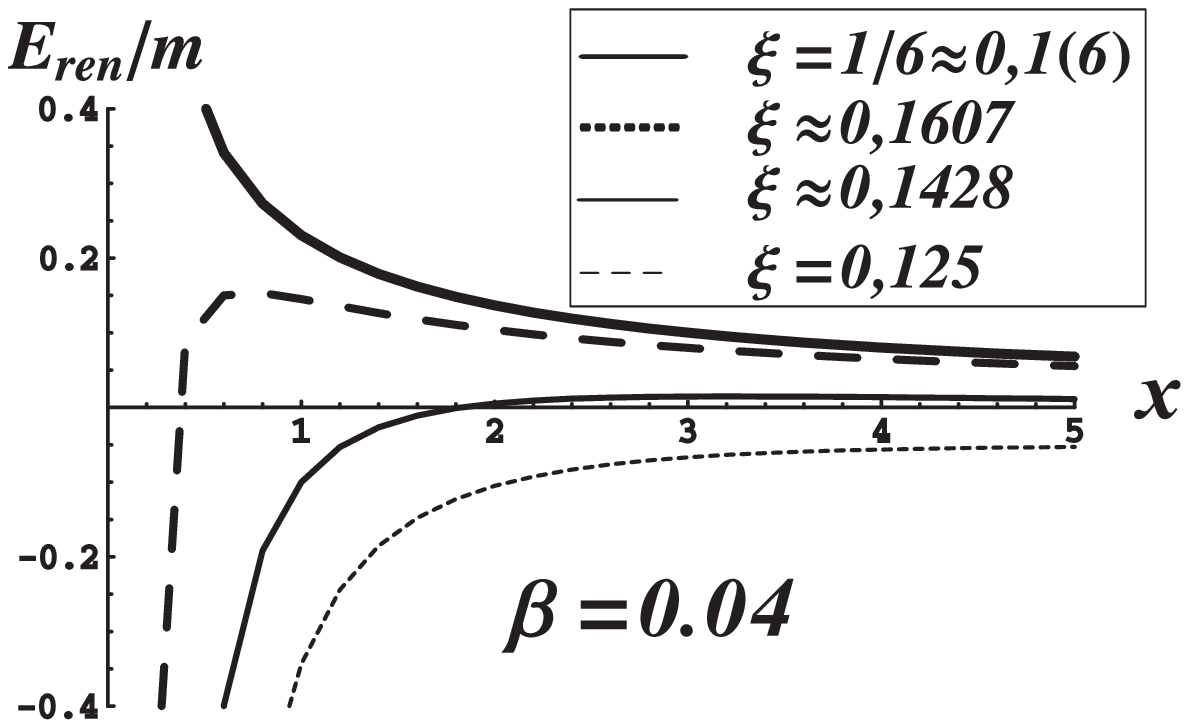}
\epsfxsize=5.5truecm\epsfbox{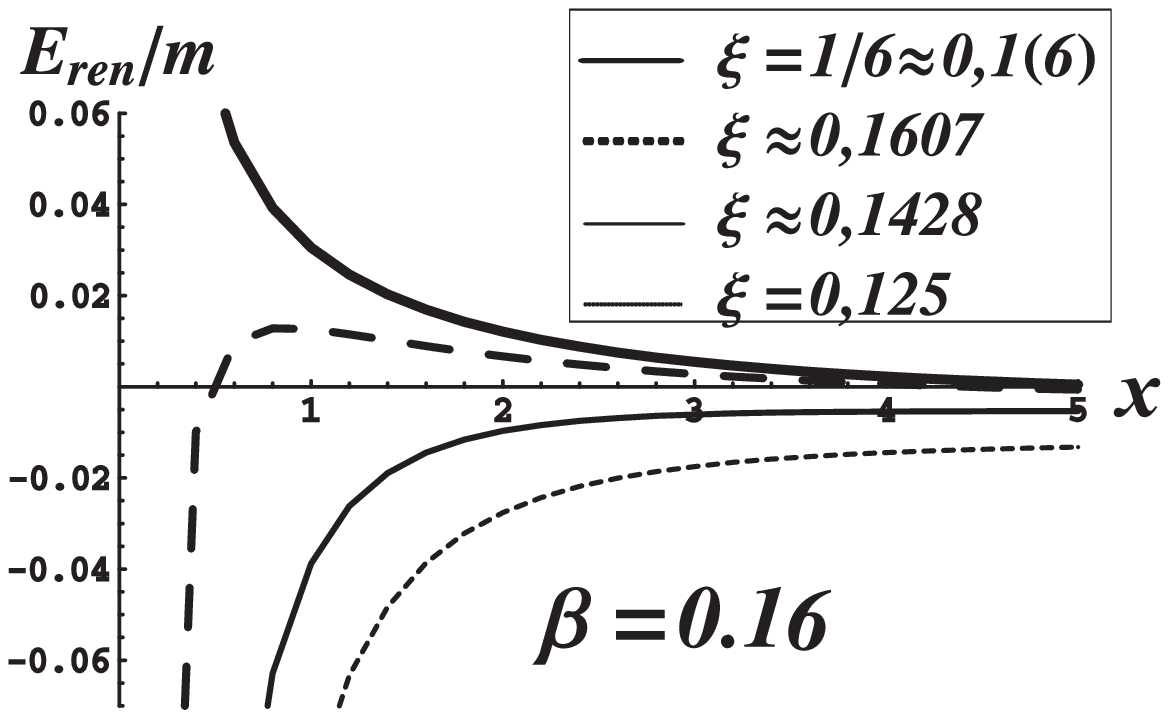}
\epsfxsize=5.5truecm\epsfbox{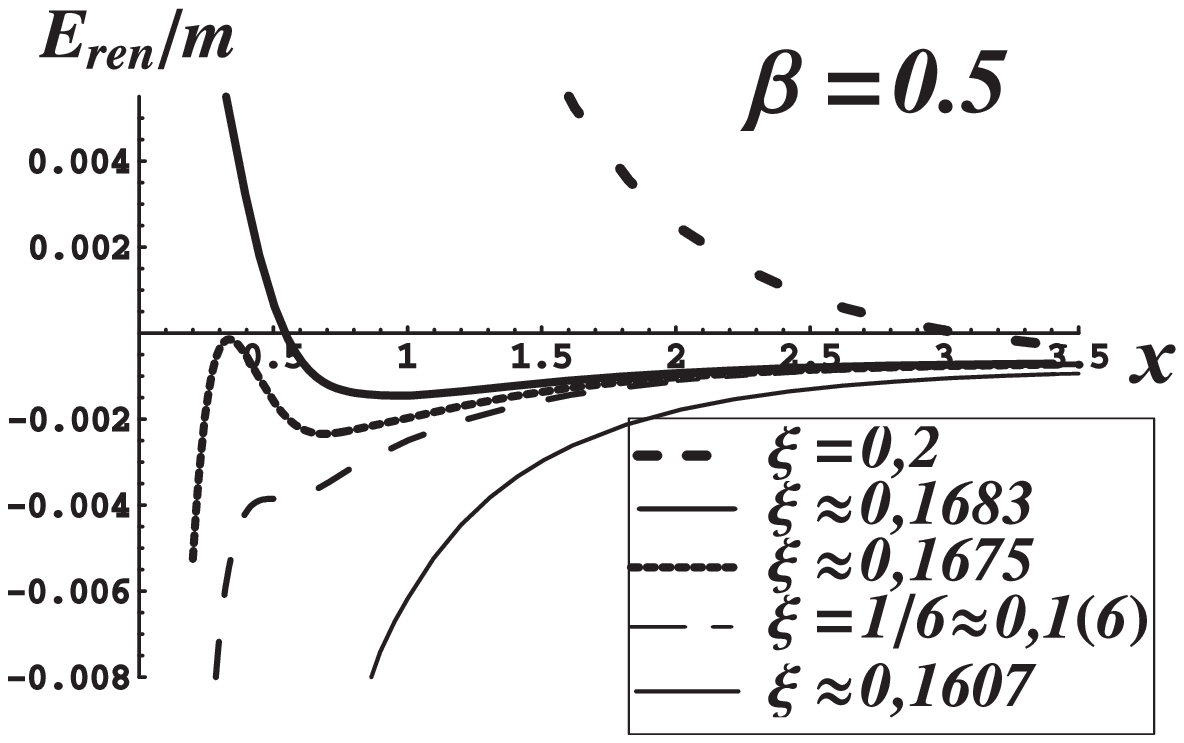}
\end{center} \caption{The plots of renormalized zero-point energy
$E_{ren}/m$ as a function of $x=R/a$ for $\beta = 0.04,0.16,0.5$
and for various values of $\xi$ and fixed mass $m$. We observe
that increasing $\xi$ leads to appearance maximum and/or minimum.
For subsequent increasing $\xi$ the curve will turn over and
extremum disappears. If the radius of spherical shell exceeds ten
radius of throat the zero-point energy takes on a value which
equals to zero-point energy in whole wormhole spacetime.}
\label{energy004}
\end{figure}

\begin{figure}
\begin{center}
\epsfxsize=5.5truecm\epsfbox{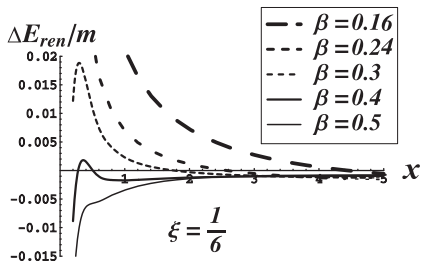}
\end{center} \caption{The plot of $\triangle E_{ren}/m$
as a function of $x=R/a$ for $\xi = \fr 16$ and for various values
of $\beta$ and fixed mass $m$. We observe the dynamics of
deformation of energy due to changing the parameter $\beta =ma$
for fixed $\xi$.} \label{energyxi}
\end{figure}

\section{Discussion and conclusion}\label{conclusion}

In this section we will discuss results of numerical calculations
of zero-point energy given by formula \Ref{dE}. The renormalized
zero-point energy is represented in figures \ref{energy004},
\ref{energyxi} as a function of $x=R/a$ (the position of sphere
rounding the wormhole) for various values of $\beta=ma$ and $\xi$.
(Note that the value $x=R/a$ characterizes the position of sphere
rounding the wormhole; $x=0$ corresponds to sphere's radius equals
to throat's radius.) In Fig. \ref{energy004} we only show the full
energy $E$. Note that the $\triangle E$ differs just slightly from
the full energy $E$. For the same reason we reproduce in Fig.
\ref{energyxi} the $\triangle E$, only.

Characterizing the result of calculations we should first of all
stress that the value of zero point energy $E_{ren}$ in the limit
$R\to\infty$ tends to some constant value obtained in Ref.
\cite{KhuSus02} for the case of wormhole spacetime without any
spherical shells. In the limit $R\to0$ (i.e., when the sphere
radius $a+R$ tends to the throat's radius $a$) the zero-point
energy $E_{ren}$ is infinitely decreasing for all $\beta$ and
$\xi$. This means that the Casimir force acting on the spherical
shell and corresponding to the Casimir zero point energy $E_{ren}$
is ``attractive'', i.e., it is directed inward to the wormhole's
throat, for sufficiently small values of $R$. In the interval
$0<R/a<\infty$ there are three qualitatively different cases of
behavior of $E_{ren}$ depending on values of $\beta$ and $\xi$.
Namely, (i) the zero point energy $E_{ren}$ is monotonically
increasing in the whole interval $0<R/a<\infty$. There are neither
maxima no minima in this case. Hence the Casimir force is
attractive for all positions of the spherical shell. (ii)
$E_{ren}$ is first increasing and then decreasing. A graph of the
zero point energy has the form of barrier with some maximal value
of $E_{ren}$ at $R_{1}/a$. The Casimir force is attractive for the
sphere's radius $R<R_{1}$ and repulsive for $R>R_{1}$. The value
$R=R_{1}$ corresponds to the point of unstable equilibrium. (iii)
The zero point energy $E_{ren}$ is increasing for $R/a<R_{1}/a$,
decreasing for $R_{1}/a<R/a<R_{2}/a$ and then finally increasing
for $R/a>R_{2}/a$, so that a graph of $E_{ren}$ has a maximum and
minimum. In this case the Casimir force is directed outward
provided the sphere's radius $R_{1}<R<R_{2}$, and inward provided
$R<R_{1}$ or $R>R_{2}$. Now the value $R=R_{2}$ corresponds to the
point of stable equilibrium, since the zero point energy $E_{ren}$
has here a local minimum.

It is worth noting that the Casimir force is attractive in the
whole interval $0<R/a<\infty$ for sufficiently small values of
$\xi$ and/or large values of $\beta$. Otherwise, it can be both
attractive and repulsive depending on a radius of sphere rounding
the wormhole's throat. The similar situation appears for
delta-like potential on the spherical or on the cylindrical
boundaries \cite{Sca99,Sca00}. The repulsive Casimir force was
also observed in Ref. \cite{HerSam05} for scalar field living in
the Einstein Static Universe.

The considered model let us speculate in spirit of Casimir idea
who suggested a model of electron as a charged spherical shell
\cite{Cas56}. Casimir assumed that such a configuration should be
stable due to equilibrium between the repulsive Coulomb force and
the attractive Casimir force. However, as is known, this idea does
not work in Minkowski spacetime since the Casimir force for sphere
turns out to be repulsive \cite{Boy68}. Now one can revive the
Casimir's idea by considering a spherical shell rounding the
wormhole. In this paper we have shown that the Casimir force now
can be both attractive and repulsive. Moreover, there exists
stable configurations for which the Casimir force equals to zero;
the radius of spherical shell in this case depends on the throat's
radius $a$ as well as the field's mass $m$ and coupling constant
$\xi$. Thus, one may try to realize the Casimir's idea taking a
sphere rounding a wormhole. Of course, our consideration was based
on the very simple model of wormhole spacetime. However, we
believe that main features of above consideration remain the same
for more realistic models.

\begin{acknowledgements}
The work was supported by part the Russian Foundation for Basic
Research grant N 05-02-17344.
\end{acknowledgements}

\end{document}